\documentclass[a4paper,10pt]{article}
\usepackage{latexsym}
\usepackage{amsmath}
\usepackage{amssymb}
\usepackage{amscd}

\newlength{\minitwocolumn}
\makeatletter
\@addtoreset{equation}{section}
\makeatother

\newtheorem{thm}{Theorem}
\newtheorem{prop}[thm]{Proposition}

\title{Smirnov-type integral formulae for 
correlation \\ functions of 
the bulk/boundary XXZ model \\ in the 
anti-ferromagnetic regime}

\author{Yas-Hiro Quano\thanks
{email: quanoy@suzuka-u.ac.jp}}

\date{\it Department of Medical Electronics, 
Suzuka University of Medical Science \\
      \it Kishioka-cho, Suzuka 510-0293, Japan}
\begin{document}

\maketitle
\begin{abstract}
Presented are the integral solutions to the quantum 
Knizhnik-Zamolodchikov equations for the 
correlation functions of both the bulk and boundary XXZ 
models in the anti-ferromagnetic regime. 
The difference equations can be derived from 
Smirnov-type master equations for correlation functions 
on the basis of the CTM bootstrap. Our integral solutions 
with an appropriate choice of the integral kernel reproduce 
the formulae previously obtained by using the bosonization 
of the vertex operators of the quantum affine algebra 
$U_q (\widehat{\mathfrak{sl}_2})$. 
\end{abstract}

\section{Introduction}

In this paper we address both the bulk and boundary XXZ 
models in the anti-ferromagnetic regime on the basis of 
Smirnov's axiomatic treatment of massive integrable 
models \cite{Smbk}. Smirnov found three axioms that 
the form factors of the sine-Gordon model should 
satisfy \cite{Smbk}. The said three axioms consist 
of {\it the $S$-matrix symmetry}, 
{\it the cyclicity condition}, and 
{\it the annihilation pole condition}. 
Smirnov also pointed out in \cite{Sm1} 
that the first two axioms imply the quantum 
KZ equations \cite{FR} of level $0$ for form factors. 

We have a similar story for correlation functions 
under the scenario based on the CTM (corner transfer 
matrix) bootstrap approach \cite{ESM}. 
As was shown in \cite{JMN} that correlation 
functions of bulk massive integrable models such as 
the XYZ model should satisfy the three relations: 
{\it the $R$-matrix symmetry}, 
{\it the cyclicity condition} and 
{\it the normalization of correlation functions}. 
Let us call those three relations 
{\it Smirnov-type master equations}. 
The first two relations imply the quantum 
Knizhnik-Zamolodchikov (KZ) equations \cite{FR} of 
level $-4$. The boundary analogue of Smirnov-type 
master equations were also obtained in \cite{JKKMW} 
for the boundary XYZ model. 
In the boundary model case, the second cyclicity condition 
should be replaced by {\it reflection properties}. 

We start from Smirnov-type master equations for 
correlation functions of the bulk/boundary XXZ model in 
the antiferromagnetic regime. We will make the Ansatz in 
the subsequent sections that the correlation functions 
are expressed in terms of integral transform of certain 
rational functions with appropriate integral kernels. 
We determine the said rational functions such that 
the $R$-matrix symmetry holds. We also find the 
transformation properties and recursion relations for 
the integral kernels by imposing the cyclicity conditions 
(resp. reflection properties) and the normalization 
conditions for the bulk (resp. boundary) XXZ model. 

The above mentioned Ansatz was already made in 
\cite{JKMQ,KMQ,affine}, where we constructed the 
form factors of the anti-ferromagnetic XXZ model 
by solving Smirnov's three axioms. We will see below 
that the integral formulae for correlation functions 
and those for form factors have the very similar structure. 
The difference between those formulae mainly results 
from the choice of rational functions, while the integral 
kernels are essentially the same. For the case of form 
factors, the freedom of solutions to the quantum KZ equation 
corresponds to that of local operators of the XXZ model. 

In the framework of the representation theory of 
the affine quantum algebra \cite{JMbk} 
the space of states of the anti-ferromagnetic XXZ model can 
be identified with the tensor product of irreducible infinite 
dimensional modules, one is level $1$ highest and the other 
is level $-1$ lowest modules of 
$U_q \bigl( \widehat{\mathfrak{sl}_2}\bigr)$ with 
$0<-q<1$. Here the restriction $0<-q<1$ is required by the 
anti-ferromagnetic condition. The correlation functions of 
the anti-ferromagnetic XXZ model are given by the trace of 
the product of the type I vertex operators on the space of 
physical states \cite{JMbk}. Furthermore, 
the integral formulae for the 
correlation functions of the XXZ model was also obtained 
in \cite{CORR} by using the bosonization of vertex operators 
of $U_q (\widehat{\mathfrak{sl}_2})$. 
The vertex operator approach were also applied to obtain
the integral formulae for correlation functions of the 
anti-ferromagnetic XXZ model with a boundary \cite{JKKKM}. 

One of advantages of the vertex operator approach is 
that correlation functions in this framework naturally 
satisfy the quantum KZ equations by the 
construction. Another advantage is that it is 
also relevant to massless models. 
In \cite{massless-XXZ} the integral formulae were 
conjectured by solving the quantum KZ equations of level 
$-4$ for correlation functions of the massless XXZ 
model, which has $U_q (\widehat{\mathfrak{sl}_2})$-symmetry 
with $|q|=1$. This was based on the CTM bootstrap approach 
\cite{ESM}, which is not directly applicable to the massless 
regime. In \cite{JKM} those conjectured formulae 
were reproduced by using the bosonized vertex operators. 
The same bosonization method was used to 
obtain the corresponding formulae for the massless XXZ model 
with a boundary in \cite{b-massless-XXZ}. 

From the view point of solving the quantum KZ equations, 
the trace construction allows us to chose the level 
arbitrarily, but gives only a canonical solution. 
On the other hand, Smirnov-type integral 
formulae based on the form factor bootstrap approach have 
the freedom of solutions but only solve the level $0$ quantum 
KZ equations. As for correlation functions, Smirnov-type 
integral formulae give solutions to the level $-4$ equations. 
The relations among those different type integral formulae 
were discussed by Nakayashiki et al \cite{NPT}. Furthermore, 
Nakayashiki proved that the trace functions of type I and 
type II vertex operators associated with the 
$U_q (\widehat{\mathfrak{sl}_n})$-symmetric model 
give a basis of the solution space of the quantum KZ equation 
at arbitrary level \cite{N1}. The CTM bootstrap approach 
was developed for $A^{(1)}_{n-1}$-symmetric elliptic models 
without and with a boundary, in order to derive the quantum 
KZ equations of level $-2n$ \cite{SPn,bBela,bJMO}. 

The rest of the present paper is organized as follows. 
In section 2 we formulate the bulk XXZ model in the 
anti-ferromagnetic regime, and present Smirnov-type integral 
formulae for correlation functions. In section 3 we prove 
that the formulae given in section 2 actually solve the quantum 
KZ equations. In section 4 we further present Smirnov-type 
integral formulae for correlation functions of the boundary 
XXZ model. In section 5 we give some concluding remarks. 
In Appendix A we give a simple proof of Proposition 
\ref{prop:b-qKZ}. 

\section{Integral formulae for the bulk XXZ model}

\subsection{The Hamiltonian and the $R$-matrix} 

In this section we consider the XXZ spin chain 
in an infinite lattice 
\begin{equation}
H_{XXZ}=-\frac{1}{2}\sum_{j\in\mathbb{Z}} 
(\sigma_{j+1}^x \sigma_{j}^x + \sigma_{j+1}^y 
\sigma_{j}^y +\Delta \sigma_{j+1}^z \sigma_{j}^z). 
\label{eq:XXZ-H}
\end{equation}
Here $\sigma_{j}^x$, $\sigma_{j}^y$ and $\sigma_{j}^z$ 
denote the standard Pauli matrices acting on $j$-th site, 
and we 
restrict ourselves to the anti-ferromagnetic regime: 
$\Delta <-1$. 
The XXZ Hamiltonian $H_{XXZ}$ commutes with 
$U_q \bigl( \widehat{\mathfrak{sl}_2}\bigr)$, where 
$\Delta=(q+q^{-1})/2$ and $-1<q<0$. For lator convenience 
we also introduce the positive parameter $x=-q$ such that 
$0<x<1$. Let $V=\mathbb{C}v_+ +\mathbb{C}v_-$ be a vector 
representation 
of $U_q \bigl( \widehat{\mathfrak{sl}_2}\bigr)$. Then 
the Hamiltonian (\ref{eq:XXZ-H}) formally acts on 
$V^{\otimes \infty}=\cdots \otimes V\otimes V\otimes 
\cdots$. In \cite{JMbk} the space of states 
$V^{\otimes \infty}$ was identified with the tensor 
product of level $1$ highest and level $-1$ lowest 
representations of 
$U_q \bigl( \widehat{\mathfrak{sl}_2}\bigr)$. 

Let us introduce the $R$-matrix of the six vertex 
model, where $R(\zeta) \in \mbox{End}(V \otimes V)$: 
\begin{equation}
R(\zeta)v_{\varepsilon_1}\otimes v_{\varepsilon_2}=
\sum_{\varepsilon'_1,\varepsilon'_2 =\pm} 
v_{\varepsilon'_1}\otimes v_{\varepsilon'_2}
R(\zeta)_{\varepsilon_1 \varepsilon_2}
    ^{\varepsilon'_1 \varepsilon'_2}, ~~~~
    R(\zeta)=\frac{1}{\kappa (\zeta )} 
    \overline{R}(\zeta), 
\label{eq:R-comp}
\end{equation}
where 
\begin{equation}
\kappa (\zeta)=\zeta
\frac{(x^4 z;x^4)_{\infty}(x^2 z^{-1};x^4)_{\infty}}
{(x^4 z^{-1};x^4)_{\infty}(x^2 z;x^4)_{\infty}}, 
~~~~
(a;p_1,\cdots,p_n)_\infty=
\prod_{k_i\ge 0}(1-ap_1^{k_1}\cdots p_n^{k_n}), 
\label{eq:kappa}
\end{equation}
and $z=\zeta^2$. The nonzero entries are given by 
\begin{equation}
\begin{array}{l}
\overline{R}(\zeta)^{++}_{++}
=\overline{R}(\zeta)^{--}_{--}=1,\\[3mm]
\displaystyle\overline{R}(\zeta)^{+-}_{+-}
=\overline{R}(\zeta)^{-+}_{-+}=b(\zeta) 
={x(\zeta^2 -1)\over 1-x^2\zeta^2},\\[3mm]
\displaystyle \overline{R}(\zeta)^{+-}_{-+}
=\overline{R}(\zeta)^{-+}_{+-}=c(\zeta)
={(1-x^2)\zeta \over 1-x^2 \zeta^2 }. 
\end{array}
\label{eq:b,c}
\end{equation}
Assume that the spectral parameter $\zeta$ and the 
constant $x$ lie in the principal regime: 
$0<x<\zeta^{-1}<1$. 
As is well known, the XXZ Hamiltonian in the 
anti-ferromagnetic regime can be obtained from the 
transfer matrix for the six vertex model in the principal 
regime, by taking logalithmic derivative with respect 
to the spectral parameter $\zeta$. 

The main properties of the $R$-matrix 
are the Yang-Baxter equation
\begin{equation}
R_{12}(\zeta_1/\zeta_2)
R_{13}(\zeta_1/\zeta_3)
R_{23}(\zeta_2/\zeta_3)=
R_{23}(\zeta_2/\zeta_3)
R_{13}(\zeta_1/\zeta_3)
R_{12}(\zeta_1/\zeta_2), 
\label{eq:YBE}
\end{equation}
where the subscript of the $R$-matrix denotes 
the spaces on which $R$ nontrivially acts; 
the initial condition 
\begin{equation}
R(1)=P; 
\label{eq:ini}
\end{equation}
the unitarity relation
\begin{equation}
R_{12}(\zeta_1/\zeta_2)R_{21}(\zeta_2/\zeta_1)=1; 
\label{eq:uni}
\end{equation}
the $\mathbb{Z}_2$-parity
\begin{equation}
R_{12}(-\zeta)=-\sigma^x_1 R_{12}(\zeta)\sigma^z_1 ; 
\label{eq:parity}
\end{equation}
and the crossing symmetries
\begin{equation}
R_{21}^{t_1}(\zeta_2/\zeta_1)=\sigma^x_1 
    R_{12}(x^{-1}\zeta_1/\zeta_2)\sigma^x_1 , ~~~~
    R_{21}^{t_1}(\zeta_2/\zeta_1)=(i\sigma^y_1 )
    R_{12}(-x^{-1}\zeta_1/\zeta_2)(i\sigma^y_1 ). 
\label{eq:cross}
\end{equation}
The properties (\ref{eq:ini}--\ref{eq:cross}) hold 
if the normalization factor of the $R$-matrix satisfies 
the following relations: 
\begin{equation}
\kappa (\zeta )\kappa (\zeta^{-1})=1, ~~~~
\kappa (\zeta )\kappa (\epsilon x\zeta)=\epsilon 
b(\zeta ), 
\label{eq:UC}
\end{equation}
where $\epsilon =\pm$. 
Under this normalization the partition function per 
lattice site is equal to unity in the thermodynamic 
limit \cite{ESM,JMbk}. 

\subsection{Correlation functions and difference equations}

Let us introduce the $V^{\otimes 2n}$-valued 
correlation functions 
\begin{equation}
G^{(n)}_\sigma(\zeta_1,\cdots,\zeta_{2n})=
\sum_{\varepsilon_j =\pm\atop \varepsilon_1 +
\cdots +\varepsilon_{2n}=0} v_{\varepsilon_1} \otimes 
\cdots \otimes v_{\varepsilon_{2n}} 
G^{(n)}_\sigma(\zeta_1,\cdots,\zeta_{2n})
^{\varepsilon_1 \cdots \varepsilon_{2n}}, 
~~~~ (\sigma =\pm ). 
\label{eq:df-corr}
\end{equation}
Here, we restrict $G^{(n)}_\sigma(\zeta)$ 
to the `total spin-$0$' subspace of $V^{\otimes 2n}$. 
In the framework of the representation theory of 
$U_q (\widehat{\mathfrak{sl}_2})$, the correlation 
function (\ref{eq:df-corr}) gives the expectation 
value of the local operator of the form 
$$
{\cal O}=E^{(1)}_{\varepsilon_1 \varepsilon'_1} 
\cdots E^{(n)}_{\varepsilon_n \varepsilon'_n}, 
$$
where $E^{(j)}_{\varepsilon_j \varepsilon'_j}$ 
is the matrix unit on the $j$-th site, by specializing 
the spectral parameters as follows: 
\begin{equation}
\langle {\cal O} \rangle_\sigma 
=G^{(n)}_\sigma(\overbrace{x^{-1}\zeta,\cdots,
x^{-1}\zeta}^{n}, \overbrace{\zeta,\cdots,\zeta}^{n})^{
-\varepsilon_{n}\cdots -\varepsilon_{1}\varepsilon'_1 
\cdots \varepsilon'_{n}}. 
\label{eq:spec}
\end{equation}

In what follows we often use the abbreviations: 
$(\zeta) =(\zeta_1 , \cdots , \zeta_{2n})$, 
$(\zeta') =(\zeta_1 , \cdots , \zeta_{2n-1})$, 
$(\zeta'') =(\zeta_1 , \cdots , \zeta_{2n-2})$, 
$(z)=(z_1 , \cdots z_{2n})$, 
$(z')=(z_1 , \cdots z_{2n-1})$; and 
$(\varepsilon )=(\varepsilon_1 \cdots \varepsilon_{2n})$, 
$(\varepsilon')=(\varepsilon_1 \cdots \varepsilon_{2n-1})$, 
$(\varepsilon'')=(\varepsilon_1 \cdots \varepsilon_{2n-2})$. 
On the basis of the CTM (corner transfer matrix) bootstrap 
approach, the correlation functions satisfy the following 
three conditions \cite{JMN}: 

\noindent{\it 1. $R$-matrix symmetry}
\begin{equation}
P_{j\,j+1} G_{\sigma}^{(n)} 
(\cdots,\zeta_{j+1},\zeta_j,\cdots) 
\quad =
R_{j\,j+1}(\zeta_j/\zeta_{j+1})G_{\sigma}^{(n)} 
(\cdots,\zeta_j,\zeta_{j+1},\cdots)
\qquad (1\leqslant j\leqslant 2n-1), 
\label{eq:R-symm} 
\end{equation}
where $P(x\otimes y)=y\otimes x$. 

\noindent{\it 2. Cyclicity}
\begin{equation}
P_{12}\cdots P_{2n-1 2n} 
G_{\sigma}^{(n)} (\zeta', x^{2}\zeta_{2n})
=\sigma G_{\sigma}^{(n)} (\zeta_{2n}, \zeta'). 
\label{eq:cyc}
\end{equation}

\noindent{\it 3. Normalization}
\begin{equation}
\begin{array}{rcl}
G_{\sigma}^{(n)} 
(\zeta'',\zeta_{2n-1}, \zeta_{2n})|_{
\zeta_{2n}=\epsilon x^{-1}\zeta_{2n-1}} &=&
G_{\epsilon\sigma}^{(n-1)} (\zeta'')
\otimes u_\epsilon  ~~~~ (\epsilon =\pm ), 
\end{array}
\label{eq:rec-G}
\end{equation}
where $u_\epsilon =v_+ \otimes v_- +
\epsilon v_- \otimes v_+$. 

These three conditions can be componentwisely recast 
as follows: 
\begin{equation}
\begin{array}{cl}
&G_{\sigma}^{(n)} 
(\cdots,\zeta_{j+1},\zeta_j,\cdots)^{\cdots 
\varepsilon_{j+1}\varepsilon_j \cdots} \\
=&\displaystyle\sum_{
\varepsilon'_j, \varepsilon'_{j+1}=\pm} 
R(\zeta_j/\zeta_{j+1})^{\varepsilon_j 
\varepsilon_{j+1}}_{\varepsilon'_j \varepsilon'_{j+1}}
G_{\sigma}^{(n)} 
(\cdots,\zeta_j,\zeta_{j+1},\cdots)^{\cdots 
\varepsilon'_{j}\varepsilon'_{j+1} \cdots}. 
\end{array}
\label{eq:R-symm-comp} 
\end{equation}
\begin{equation}
G_{\sigma}^{(n)} 
(\zeta', x^{2}\zeta_{2n})^{\varepsilon' \varepsilon_{2n}}
=\sigma G_{\sigma}^{(n)} 
(\zeta_{2n}, \zeta')^{\varepsilon_{2n}\varepsilon'}. 
\label{eq:cyc-comp}
\end{equation}
\begin{equation}
\begin{array}{rcl}
G_{\sigma}^{(n)} 
(\zeta'',\zeta_{2n-1}, \epsilon x^{-1}\zeta_{2n-1})^{
\varepsilon''\, ss'} &=&s^{(1-\epsilon )/2}
\delta_{s+s',0}
G_{\epsilon\sigma}^{(n-1)} (\zeta'')^{\varepsilon''} 
~~~~ (s,s'=\pm ). 
\end{array}
\label{eq:rec-G-comp}
\end{equation}
Combining (\ref{eq:R-symm-comp}) and (\ref{eq:rec-G-comp}) 
we obtain another expression of the normalization 
condition: 
$$
\displaystyle\sum_{s=\pm}
s^{(1-\epsilon )/2} G_{\sigma}^{(n)} 
(\zeta'',\zeta_{2n-1}, \epsilon x\zeta_{2n-1})^{
\varepsilon''\, s\, -s}=
G_{\epsilon\sigma}^{(n-1)} (\zeta'')^{\varepsilon''}, 
$$
where we also use 
$$
R(\epsilon x^{-1})=\left( \begin{array}{cccc} 
0 & 0 & 0 & 0 \\
0 & \epsilon & 1 & 0 \\ 0 & 1 & \epsilon & 0 \\ 
0 & 0 & 0 & 0 \end{array} \right). 
$$

Since the $R$-matrix preserves the `total spin' of 
the correlation function, we can restrict 
$G_{\sigma}^{(n)}(\zeta)$ to an element of the 
`total spin-$0$' subspace, when we consider the equations 
(\ref{eq:R-symm}--\ref{eq:rec-G}). Note that the first 
two conditions imply the difference equation of 
the quantum KZ type \cite{FR} of level $-4$: 
\begin{equation}
\begin{array}{l}
T_j G_{\sigma}^{(n)} (\zeta)
=
R_{j\,j-1}(x^{-2} \zeta_j/\zeta_{j-1})
\cdots
R_{j\,1}(x^{-2} \zeta_{j}/\zeta_1 ) \\
\qquad\qquad \times
R_{j\,2n}(\zeta_{j}/\zeta_{2n})
\cdots
R_{j\,j+1}(\zeta_{j}/\zeta_{j+1})
G_{\sigma}^{(n)} (\zeta), 
\label{eq:qKZ}
\end{array}
\end{equation}
where $T_j$ is the shift operator such that 
\begin{equation}
T_j F (\zeta)
=F (\zeta_1,\cdots,x^{-2}\zeta_j,\cdots,\zeta_{2n}), 
\label{eq:T-shift}
\end{equation}
for any $2n$-variate function $F$. When the definition 
of $T_j$ is replaced by 
$$
T_j F (\zeta)
=F (\zeta_1,\cdots,x^{l+2}\zeta_j,\cdots,\zeta_{2n}), 
$$
and all the arguments $x^{-2}\zeta_j /\zeta_k$ in the 
first line of the RHS of (\ref{eq:qKZ}) are also replaced 
by $x^{l+2}\zeta_j /\zeta_k$, the difference equation 
(\ref{eq:qKZ}) is called the quantum KZ equations of 
level $l$. 

{\bf Remark.} As a result of $\mathbb{Z}_2$-symmetry 
of $R$-matrix, there are two ground states for 
the anti-ferromagnetic XXZ model. Let us specify the 
two ground states by $i=0, 1$, and denote the correlation 
function on the $i$-th ground state by $G^{(n)}_i (\zeta)$. 
The CTM bootstrap approach suggest that both 
$G^{(n)}_0 (\zeta)$ and $G^{(n)}_1 (\zeta)$ will appear 
in the cyclicity condition as follows: 
$$
P_{12}\cdots P_{2n-1 2n} G_{i}^{(n)} 
(\zeta_1,\cdots,\zeta_{2n-1}, x^{2}\zeta_{2n})
=G_{1-i}^{(n)} 
(\zeta_{2n}, \zeta_1,\cdots,\zeta_{2n-1}). 
$$
Thus we introduce $G^{(n)}_\sigma(\zeta)
=G^{(n)}_0 (\zeta)+\sigma G^{(n)}_1 (\zeta)$ 
such that the second equation (\ref{eq:cyc}) 
involves only $G^{(n)}_\sigma (\zeta)$. 

\subsection{Integral formulae} 

Set 
\begin{equation}
 G_{\sigma}^{(n)}(\zeta)
=c_n \prod_{1\leqslant j< k \leqslant 2n} \zeta_j g(z_j/z_k)
\times \overline{G}_{\sigma}^{(n)}(\zeta). 
\label{eq:G-bar}
\end{equation}
Here $c_n$ is a constant which will be determined below, 
and the function $g(z)$ has the properties 
\begin{equation}
g(z)=g(x^{-4}z^{-1}), ~~~~ 
\kappa (\zeta )=\zeta \frac{g(z)}{g(z^{-1})}. 
\label{eq:g-prop}
\end{equation}
The explicit form of $g(z)$ is as follows: 
\begin{equation}
g(z)=\frac{(x^6 z;x^4,x^4)_{\infty}
(x^2 z^{-1};x^4,x^4)_{\infty}}
{(x^8 z;x^4,x^4)_{\infty}(x^4 z^{-1};x^4,x^4)_{\infty}}. 
\label{eq:df-g}
\end{equation}
Thanks to (\ref{eq:g-prop}) the first two equations 
(\ref{eq:R-symm-comp}--\ref{eq:cyc-comp}) are 
rephrased in terms of 
$\overline{G}_{\sigma}^{(n)}(\zeta)$ 
and $\overline{R}(\zeta)$ as follows 
\begin{equation}
\overline{G}_{\sigma}^{(n)} 
(\cdots,\zeta_{j+1},\zeta_j,\cdots)^{\cdots 
\varepsilon_{j+1}\varepsilon_j\cdots}
=\sum_{\varepsilon'_j , \varepsilon'_{j+1}=\pm} 
\overline{R}(\zeta_j/\zeta_{j+1})^{
\varepsilon_j , \varepsilon_{j+1}}_{
\varepsilon'_j , \varepsilon'_{j+1}}
\overline{G}_{\sigma}^{(n)} 
(\cdots,\zeta_j,\zeta_{j+1},\cdots)^{
\varepsilon'_j , \varepsilon'_{j+1}}, 
\label{eq:R-symm'} 
\end{equation}
\begin{equation}
\overline{G}_{\sigma}^{(n)} 
(\zeta', x^{2}\zeta_{2n})^{
\varepsilon'\,\varepsilon_{2n}}
=\overline{G}_{\sigma}^{(n)} 
(\zeta_{2n},\zeta')^{
\varepsilon_{2n}\,\varepsilon'}
\prod_{j=1}^{2n-1} \frac{\zeta_{2n}}{\zeta_j}.
\label{eq:cyc'}
\end{equation}

In order to present our integral formulae for 
$\overline{G}_{\sigma}^{(n)}(\zeta )$ let us 
prepare some notation. Let 
\begin{equation}
A:=\{ a|\varepsilon_a >0, \,\,1\leqslant a\leqslant 2n \}. 
\label{eq:df-A}
\end{equation}
Then the number of elements of $A$ is equal to $n$, 
because we consider only the `total-spin-$0$' case. 
We often use the abbreviations 
$(w)=(w_{a_1}, \cdots , w_{a_n})$, 
$(w')=(w_{a_1}, \cdots , w_{a_{n-1}})$ and 
$(w'')=(w_{a_1}, \cdots , w_{a_{n-2}})$ for $a_j \in A$ 
such that $a_1 <\cdots <a_n$. Let us define the 
following $V^{\otimes 2n}$-valued meromorphic function 
\begin{equation}
Q^{(n)}(w|\zeta )=\sum_{\varepsilon_j =\pm\atop 
\varepsilon_1 +\cdots +\varepsilon_{2n}=0} 
v_{\varepsilon_1}\otimes \cdots \otimes 
v_{\varepsilon_{2n}} 
Q^{(n)}(w|\zeta )^{\varepsilon}, 
\label{eq:vec-Q}
\end{equation}
where 
\begin{equation}
Q^{(n)}(w|\zeta )^{\varepsilon} 
=\prod_{a\in A} \zeta_a \left( 
\prod_{j=1}^{a-1} (xz_j -w_a )
\prod_{j=a+1}^{2n} (xw_a -z_j ) \right) \left/ 
\prod_{a,b\in A\atop a<b} (x^{-1}w_b -xw_a ). \right.
\label{eq:df-Q}
\end{equation}
Note that the $Q_n (\alpha |\beta )^{\varepsilon}$ 
in \cite{massless-XXZ}, the massless analogue of 
$Q^{(n)}(w|\zeta )^{\varepsilon}$, can be obtained 
by substituting $\zeta_j =e^{-\nu \beta_j}$, 
$x=e^{\pi i\nu}$, and $w_a =e^{-2\nu \alpha_a}$ into 
(\ref{eq:df-Q}), up to a trivial factor. 

We wish to find integral formulae of the form 
\begin{equation}
\overline{G}_{\sigma}^{(n)} 
(\zeta)=\prod_{a\in A}\oint_{C_a} 
\dfrac{dw_a}{2\pi iw_a} 
\Psi_{\sigma}^{(n)} (w|\zeta )
Q^{(n)}(w|\zeta). 
\label{eq:G-form}
\end{equation}
Here, the kernel has the form 
\begin{equation}
\Psi^{(n)}_\sigma (w| \zeta )=
\vartheta^{(n)}_\sigma (w | \zeta )
\prod_{a\in A}\prod_{j=1}^{2n} 
\psi \Bigl(\frac{w_{a}}{z_j}\Bigr),
\label{eq:df-Psi}
\end{equation}
where
\begin{equation}
\psi(z)=\frac{1}{(xz;x^4)_{\infty}
(xz^{-1};x^4)_{\infty}}. 
\label{eq:df-psi}
\end{equation}
For the function 
$\vartheta^{(n)}_\sigma(w|\zeta )$ 
we assume that
\begin{itemize}
\item
it is anti-symmetric and holomorphic
in the $w_a \in \mathbb{C} \backslash \{0\}$,
 
\item
it is symmetric and meromorphic
in the $\zeta_j\in \mathbb{C} \backslash \{0\}$,
 
\item
it has the two transformation properties 
\begin{eqnarray}
\displaystyle\frac
{\vartheta^{(n)}_\sigma (w |\zeta', x^2 \zeta_{2n})}
{\vartheta^{(n)}_\sigma (w | \zeta )}&=& 
\sigma \displaystyle\prod_{a\in A}
\frac{w_{a}}{x^4 z_{2n}} 
\prod_{j=1}^{2n-1} \frac{\zeta_{2n}}{\zeta_j}=
\sigma x^{-4n} \displaystyle\prod_{a\in A}w_{a}
\prod_{j=1}^{2n}\zeta_j^{-1}, 
\label{eq:z-sym}\\
\displaystyle \frac
{\vartheta^{(n)}_\sigma (w', x^4 w_{a_n}| \zeta )}
{\vartheta^{(n)}_\sigma (w | \zeta )}
&=& x^{2n}\displaystyle\prod_{j=1}^{2n} 
\frac{z_j }{x^4 w_{a_n}}=x^{-6n}
\displaystyle\prod_{j=1}^{2n} \frac{z_j }{w_{a_n}}.
\label{eq:x-sym}
\end{eqnarray}
\item it satisfies the following recursion relation
\begin{equation}
\displaystyle \frac{\vartheta_{\sigma}^{(n)}
(w', x^{-1}z_{2n-1} |\zeta'',\zeta_{2n-1}, 
\epsilon x^{-1} \zeta_{2n-1})}
{\vartheta_{\epsilon\sigma}^{(n-1)}(w' |\zeta'')}=
\displaystyle x^{2n}z_{2n-1}^{-2n}\prod_{j=1}^{2n-2} 
z_j^{-1} \prod_{a\in A\atop a\neq a_{2n}}w_a^{-1} 
\Theta_{x^2}(x w_{a}/z_{2n-1}). 
\label{eq:th-rec}
\end{equation}
Here 
$$
\Theta_{p}(z):=(z; p)_\infty 
(pz^{-1}; p)_\infty (p, p)_\infty =
\sum_{m\in\mathbb{Z}} p^{m(m-1)}(-z)^m , 
$$
and we also fix the constant $c_n$ as follows: 
\begin{equation}
c_n =\frac{(-1)^{n(n+1)/2}}{x^{n(n-1)/2}
(x^2 ; x^2 )_\infty^{n(n-1)/2}}\frac{
(x^2 ; x^4 , x^4 )_\infty^{2n}}{
(x^8 ; x^4 , x^4 )_\infty^{2n}}. 
\label{eq:c_n}
\end{equation}
\end{itemize}
The function $\vartheta_{\sigma}^{(n)}(w|\zeta )$
is otherwise arbitrary, and the choice of 
$\vartheta_{\sigma}^{(n)}(w|\zeta )$ 
corresponds to that of solutions to 
(\ref{eq:R-symm}--\ref{eq:rec-G}). 
The transformation property of 
$\vartheta_{\sigma}^{(n)}(w|\zeta )$ implies
\begin{eqnarray}
\frac{\Psi_{\sigma}^{(n)}(w|\zeta', x^2 \zeta_{2n})}
{\Psi_{\sigma}^{(n)}(w|\zeta )}
&=&\sigma\prod_{a\in A}\frac{w_a-xz_{2n}}{x^4 z_{2n}-xw_a}
\prod_{j=1}^{2n} \frac{\zeta_{2n}}{\zeta_j}, 
\label{eq:trPsi1} \\
\frac{\Psi_{\sigma}^{(n)}(w',x^4 w_{a_n}|\zeta )}
{\Psi_{\sigma}^{(n)}(w|\zeta )}
&=&\prod_{j=1}^{2n}\frac{z_j -xw_{a_n}}{x^3 w_{a_n}-z_j}.
\label{eq:trPsi2}
\end{eqnarray}

The integrand may have poles at 
$$
w_a =\left\{ 
\begin{array}{ll}
x^{\pm (1+4k)}z_j & (1\leqslant j\leqslant 2n, 
k\in \mathbb{Z}_{\geqslant 0}), \\
x^2 w_b & (b<a), \\
x^{-2} w_b & (b>a). 
\end{array} \right. 
$$
We choose the integration contour $C_a$ with 
respect to $w_a$ ($a\in A$) such that $C_a$ is along 
a simple closed curve oriented anti-clockwise, and 
encircles the points $x^{1+4k}z_j 
(1\leqslant j \leqslant 2n , k\in \mathbb{Z}_{\geqslant 0})$ 
and $x^2 w_b$ ($b<a$) but not $x^{-1-4k} z_j 
(1\leqslant j \leqslant 2n, k\in \mathbb{Z}_{\geqslant 0})$ 
nor $x^{-2} w_b$ ($b>a$). Thus the contour $C_a$ actually 
depends on $z_j$'s besides $a$, so that it precisely should 
be denoted by $C_a (z)=C_a (z_1, \cdots , z_{2n})$. The LHS 
of (\ref{eq:cyc}) refers to the analytic continuation with 
respect to $\zeta_{2n}$. 
Nevertheless, once we restrict ourselves 
to the principal regime $0<x<\zeta_j^{-1}<1$, 
we can tune all $C_a$'s to be the common integration 
contour $C: |w_a|=x^{-1}$ because of the inequality 
$xz_j <x^{-1}<x^{-1}z_j$. 

\section{Main proposition}

In this section we prove the following proposition in 
the bulk anti-ferromagnetic XXZ model case: 

\begin{prop} Assume the properties of 
the function $\vartheta^{(n)}_\sigma$ 
below (\ref{eq:df-psi}) and the integration contour 
$C_a$ below (\ref{eq:c_n}). Then the integral formulae 
(\ref{eq:df-corr}, \ref{eq:G-bar}, \ref{eq:G-form}) 
with (\ref{eq:df-g}, \ref{eq:vec-Q}, \ref{eq:df-Q}, 
\ref{eq:df-Psi}, \ref{eq:df-psi}) 
solves the three equations 
(\ref{eq:R-symm}--\ref{eq:rec-G}). 
\label{prop:qKZ}
\end{prop}

\subsection{Proof of the $R$-matrix symmetry} 

Let us first prove (\ref{eq:R-symm}), or equivalently 
(\ref{eq:R-symm'}), in a componentwise way. 

Suppose that $j,j+1\not\in A$. Then the relation 
(\ref{eq:R-symm'}) holds, because the integrand 
$\overline{G}^{(n)}_\sigma (\zeta )$ is 
symmetric with respect to $\zeta_j$ and $\zeta_{j+1}$. 

When $j\not\in A$ and $j+1\in A$ the relation 
(\ref{eq:R-symm'}) reduces to 
\begin{equation}
\begin{array}{cl}
&\overline{G}_{\sigma}^{(n)} 
(\cdots,\zeta_{j+1},\zeta_j,\cdots)^{\cdots 
+- \cdots} \\
=&b(\zeta_j/\zeta_{j+1})
\overline{G}_{\sigma}^{(n)} 
(\cdots,\zeta_j,\zeta_{j+1},\cdots)^{\cdots 
-+ \cdots}+c(\zeta_j/\zeta_{j+1})
\overline{G}_{\sigma}^{(n)} 
(\cdots,\zeta_j,\zeta_{j+1},\cdots)^{\cdots 
+- \cdots}. 
\end{array}
\label{eq:R-symm/-+}
\end{equation}
Note that the set of the integration variables in 
the second term of the RHS is different from the 
other terms. Since $w_a$'s are integration variables, 
we can replace $w_{j}$ in the second term of the RHS 
by $w_{j+1}$. After that, the relation (\ref{eq:R-symm/-+}) 
follows from the equality of the integrands. In this step 
we use 
$$
xw_{j+1}-z_j =\frac{x(z_j -z_{j+1})}{z_{j+1}-x^2 z_j}
(xz_j -w_{j+1})+\frac{(1-x^2 )z_j}{z_{j+1}-x^2 z_j}
(xw_{j+1} -z_{j+1}). 
$$

The case $j\in A$ and $j+1\not\in A$ can be proved in 
a similar way to the previous case. Here we use 
$$
xz_{j+1}-w_j =\frac{x(z_j -z_{j+1})}{z_{j+1}-x^2 z_j}
(xw_j -z_{j+1})+\frac{(1-x^2 )z_{j+1}}{z_{j+1}-x^2 z_j}
(xz_{j} -w_{j}). 
$$

Finally consider the case $j, j+1\in A$. Since 
$\Psi^{(n)}_\sigma (w|\zeta )$ is antisymmetric 
with respect to $w_a$'s, only the antisymmetric part of 
$Q^{(n)}(w|\zeta )^{\cdots ++\cdots}$ with respect to 
$w_j$ and $w_{j+1}$ gives non-zero contribution for 
the integral. Thus the relation (\ref{eq:R-symm'}) 
in this case follows from the fact that the 
following combination 
$$
\frac{(xw_j -z_{j+1})(xz_j -w_{j+1})}
{x^{-1}w_{j+1}-xw_j}-
\frac{(xw_{j+1} -z_{j+1})(xz_j -w_{j})}
{x^{-1}w_{j}-xw_{j+1}}
$$
is symmetric with respect to $z_j$ and $z_{j+1}$. 

\subsection{Proof of the cyclicity} 

Let $2n\not\in A$. When the integral (\ref{eq:G-form}) 
is analytically continued from $\zeta_{2n}$ to 
$x^2 \zeta_{2n}$, the points 
$$
\cdots\cdots, \, x^9 z_{2n}, \, 
x^5 z_{2n}, \, xz_{2n}, \, x^{-1}z_{2n}, 
\, x^{-5}z_{2n}, \, x^{-9}z_{2n}, \, \cdots\cdots
$$ 
move to the points 
$$
\cdots\cdots, \, x^{13}z_{2n}, \, 
x^9 z_{2n}, \, x^5 z_{2n}, \, x^3 z_{2n}, 
\, x^{-1}z_{2n}, \, x^{-5}z_{2n}, \, \cdots\cdots. 
$$ 
In the LHS of (\ref{eq:cyc'}) the point $x^3 z_{2n}$ 
and consequently $xz_{2n}$ are outside the integral 
contour 
$C'_a=C_a (z', x^4 z_{2n})$. 
Nevertheless, we can deform $C'_a$ to the original one 
$C_a =C_a (z)$ 
without crossing any poles. That is because 
the factor 
$$
\prod_{a\in A} (xw_a -x^4 z_{2n}), 
$$
contained in 
$Q^{(n)}(w|\zeta' , x^2 \zeta_{2n})^{\varepsilon'\,-}$ 
cancels the singularity at $w_a =x^3 z_{2n}$. 
Thus the integral contours for both sides of 
(\ref{eq:cyc'}) coincide. 

Furthermore, by using (\ref{eq:trPsi1}) we obtain 
$$
\Psi^{(n)}_\sigma (w|\zeta', x^2 \zeta_{2n})
\prod_{a\in A} (xw_a -x^4 z_{2n})=\sigma
\Psi^{(n)}_\sigma (w|\zeta)
\prod_{a\in A} (xz_{2n}-w_a ) 
\prod_{j=1}^{2n} \frac{\zeta_{2n}}{\zeta_j}, 
$$
which implies that the integrands of both sides of 
(\ref{eq:cyc'}) coincide and therefore the relation 
(\ref{eq:cyc'}) holds when $\varepsilon_{2n}<0$. 

Next let $2n\in A$. In this case we also make the rescale 
of variable $w_{2n}\mapsto x^4 w_{2n}$ in the LHS of 
(\ref{eq:cyc'}). Then the integral 
contur with respect to $w_a$ ($a\in A\backslash \{ 2n\}$) 
will be $C'_a =C_a (z', x^4 z_{2n})$, 
and the other one will be 
$\widetilde{C}=C_{2n}(x^{-4}z', z_{2n})$. 
For $a\in A\backslash \{ 2n\}$, we can deform the contour 
$C'_a$ to the original $C_a$ without crossing any poles, 
as the same reason as before. The integral contour 
$\widetilde{C}$ encircles $x^{-3+4k}z_j$ 
($1\leqslant j\leqslant 2n-1$, 
$k\in \mathbb{Z}_{\geqslant 0}$) and $x^{1+4k}z_{2n}$ 
($k\in \mathbb{Z}_{\geqslant 0}$), but not 
$x^{-5-4k}z_j$ ($1\leqslant j\leqslant 2n-1$, 
$k\in \mathbb{Z}_{\geqslant 0}$) nor $x^{-1-4k}z_{2n}$ 
($k\in \mathbb{Z}_{\geqslant 0}$). Since 
$Q^{(n)}(w|\zeta' , x^2 \zeta_{2n})^{\varepsilon'\,+}$ 
contains the factor 
\begin{equation}
x^2 \zeta_{2n}\prod_{j=1}^{2n-1} (xz_j -x^4 w_{2n})
\prod_{a\in A\atop a\neq 2n} 
\frac{xw_a -x^4 z_{2n}}{x^{-1}x^4w_{2n}-xw_a}, 
\label{eq:Qtr}
\end{equation}
the pole at $w_{2n}=x^{-3}z_{2n}$ ($1\leqslant j\leqslant 2n-1$) 
disappears. Thus we can deform the contour 
$\widetilde{C}$ to the original one $C_{2n}=C_{2n}(z)$ 
without crossing any poles. Thus the integral contours 
in both sides of (\ref{eq:cyc'}) coincide. 

Replace the integral variables such that 
$(w', w_{2n})\mapsto (w_{2n}, w')$ in the RHS 
of (\ref{eq:cyc'}), and compare the integrands of both 
sides. From (\ref{eq:trPsi1}, \ref{eq:trPsi2}) and the 
antisymmetric property of $\Psi^{(n)}_\sigma$ 
with respect to $w_a$'s, we have 
\begin{equation}
\frac{\Psi^{(n)}_\sigma (w', x^4 w_{2n}|
\zeta', x^2 \zeta_{2n})}{\Psi^{(n)}_\sigma 
(w_{2n}, w'|\zeta )}=(-1)^{n-1}
x^{-3} \prod_{a\in A\atop a\neq 2n} 
\frac{w_a -xz_{2n}}{x^4 z_{2n}-xw_a}
\prod_{j=1}^{2n-1} \frac{\zeta_{2n}}{\zeta_j} 
\frac{z_j -xw_{2n}}{x^3 w_{2n}-z_j}. 
\label{eq:Psi12}
\end{equation}
Note that (\ref{eq:Qtr}) describes all $w_{2n}$- and 
$z_{2n}$-dependence of 
$Q^{(n)}(w|\zeta', x^2 \zeta_{2n})^{\varepsilon\,+}$. 
The product of (\ref{eq:Qtr}) and (\ref{eq:Psi12}) 
is equal to 
$$
\zeta_{2n} 
\prod_{j=1}^{2n-1} \frac{\zeta_{2n}}{\zeta_j} 
(xw_{2n}-z_j )\prod_{a\in A\atop a\neq 2n} 
\frac{xz_{2n}-w_a}{x^{-1}w_a -xw_{2n}}, 
$$
which implies that the integrands of both sides of 
(\ref{eq:cyc'}) coincide and therefore the relation 
(\ref{eq:cyc'}) holds when $\varepsilon_{2n}>0$. 

\subsection{Proof of the normalization condition} 

Let us prove (\ref{eq:rec-G}), or equivalently 
(\ref{eq:rec-G-comp}). 
The factor $g(z_{2n-1}/z_{2n})$ hase a zero at 
$\zeta_{2n}=\epsilon x^{-1}\zeta_{2n-1}$. 
On the other hand, 
the two points $xz_{2n}$ and $x^{-1}z_{2n-1}$ are 
required to locate opposite sides of the integral 
contour $C_a$, so that a pinching may occur as 
$\zeta_{2n}\rightarrow \epsilon x^{-1}\zeta_{2n-1}$. 
If there is no pinching the correlation function 
$G^{(n)}_\sigma (\zeta )$ vanishes at 
$\zeta_{2n}=\epsilon x^{-1}\zeta_{2n-1}$. 

Suppose $2n-1, 2n\not\in A$. In this case 
the factor $(xw_a -z_{2n-1})$ contained in 
$Q^{(n)}(w|\zeta)^{\varepsilon''\,--}$ cancel the poles of 
$\psi (w_a /z_{2n-1})$ at $w_a =x^{-1}z_{2n-1}$, 
and therefore no pinching occurs as 
$\zeta_{2n}\rightarrow x^{-1}\zeta_{2n-1}$. 
Thus the condition (\ref{eq:rec-G-comp}) holds 
if $2n-1, 2n\not\in A$. 

When $2n-1, 2n\in A$, no pinching occurs for the 
integral contour for $C_a$ ($a\in A\backslash 
\{ 2n-1, 2n\}$), as the same reason as before. 
Concerning the integral with respect to $w_{2n-1}$ 
and $w_{2n}$, there is no singularities 
at $w_{2n-1}=xz_{2n}$ and $w_{2n}=xz_{2n}$ and 
consequently no pinching occurs actually. 
In order to see such vanishing singularities, 
let us first evaluate the residue at $w_{2n}=xz_{2n}$. 
Because of the antisymmetric property of 
$\vartheta^{(n)}_\sigma$ with respect to $w_a$'s, 
the zero of 
$\vartheta^{(n)}_\sigma (w'', w_{2n-1}, xz_{2n}|\zeta )$ 
at $w_{2n-1}=xz_{2n}$ cancels the poles of 
$\psi (w_{2n-1}/z_{2n})$ at the same point. 
Furthermore, the poles of $\psi (w_{2n}/z_{2n})$ 
and $\psi (w_{2n}/z_{2n-1})$ at $w_{2n}=xz_{2n}$ 
and $z_{2n}=x^{-2}z_{2n-1}$ are canceled by 
the zeros of $g(z_{2n-1}/z_{2n})$ and 
$Q^{(n)}(w'', w_{2n-1}, xz_{2n}|\zeta )^{\varepsilon''\,++}$ 
at the same points. The latter cancellation can be shown 
as follows. 
Note that the integral is invariant as we replace 
$Q^{(n)}(w'', w_{2n-1}, xz_{2n}|\zeta )^{
\varepsilon''\,++}$ 
by its antisymmetric part with respect to 
$w_{2n-1}$ and $xz_{2n}$. When $z_{2n}=x^{-2}z_{2n-1}$ 
the antisymmetric part contains the following 
vanishing factor: 
$$
\frac{(xw_{2n-1}-z_{2n})(xz_{2n-1}-xz_{2n})}
{x^{-1}xz_{2n}-xw_{2n-1}}-
\frac{(xxz_{2n}-z_{2n})(xz_{2n-1}-w_{2n-1})}
{x^{-1}w_{2n-1}-xxz_{2n}}. 
$$
Thus there is no singularity at $w_{2n}=xz_{2n}$ as 
$\zeta_{2n}\rightarrow \epsilon x^{-1}\zeta_{2n-1}$. 
The same thing at $w_{2n-1}=xz_{2n}$ can be easily shown 
in the same way. Hence the condition (\ref{eq:rec-G-comp}) 
is verified when $2n-1, 2n\in A$. 

Next let $2n-1 \not\in A$ and $2n \in A$. 
In this case there is no pinching for the integrals 
with respect to $w_a$ ($a\in A\backslash \{ 2n \}$). 
Let $\widehat{C}$ denote the integral contour with 
respect to $w_{2n}$ such that $\widehat{C}$ encircles 
the same points as $C_{2n}$ does but $xz_{2n}$. Note that 
no pinching occurs with respect to the integral along 
$\widehat{C}$ because both the two points $x^{-1}z_{2n-1}$ 
and $xz_{2n}$ lie outside the contour $\widehat{C}$. 
Thus the integral with respect to $w_{2n}$ along 
the contour $C_{2n}$ can be replaced by the 
residue at $w_{2n}=xz_{2n}$. 

In order to evaluate the residue, the following formulae 
are useful: 
\begin{equation}
\lim_{z_{2n}\rightarrow x^{-2}z_{2n-1}} 
g\left( \frac{z_{2n-1}}{z_{2n}} \right) 
\psi \left( \frac{xz_{2n}}{z_{2n-1}} \right) 
= \frac{(x^6; x^4)_\infty}{(x^4; x^4)_\infty}
\frac{(x^4; x^4 , x^4)^2_\infty}
{(x^2; x^4 , x^4)^2_\infty}, 
\label{eq:lim_z}
\end{equation}
\begin{equation}
{\rm Res}_{w_{2n}=xz_{2n}}\frac{dw_{2n}}{w_{2n}}
\psi \left( \frac{w_{2n}}{z_{2n}} \right) 
=\frac{1}{(x^2; x^2 )_\infty}. 
\label{eq:Res_w}
\end{equation}
\begin{equation}
g(z)g(x^2 z)\psi (xz)=\frac{1}{1-x^2 z}. 
\label{eq:ggpsi_j}
\end{equation}
\begin{equation}
\psi \left( \frac{w}{z} \right) 
\psi \left( \frac{x^2 w}{z} \right) 
(z-xw) =\frac{-xw}
{(xw/z; x^2)_\infty(xz/w; x^2)_\infty}. 
\label{eq:ppQ_a}
\end{equation}
Using these the condition (\ref{eq:rec-G-comp}) 
with $\varepsilon_{2n-1}<0$ and 
$\varepsilon_{2n}>0$ reduces to 
$$
\begin{array}{rcl}
c_{n-1} \vartheta_{\epsilon\sigma}^{(n-1)}(w'|\zeta'')
&=&c_n \vartheta_\sigma^{(n)}(w', x^{-1}z_{2n-1}|
\zeta'', \zeta_{2n-1}, \epsilon x^{-1}\zeta_{2n-1}) \\
&\times& \displaystyle (-1)^n x^{-n-1} 
\frac{(x^8; x^4 , x^4)^2_\infty}
{(x^2; x^4 , x^4)^2_\infty} z_{2n-1}^{2n} 
\prod_{j=1}^{2n-2} z_j 
\prod_{a\in A\atop a\neq 2n} 
\frac{w_a (x^2; x^2 )_\infty}
{\Theta_{x^2}(xw_a /z_{2n-1})}, 
\end{array}
$$
which is valid under the assumption of 
(\ref{eq:th-rec}) and ({\ref{eq:c_n}). 

When $2n-1\in A$ and $2n\not\in A$, the only difference 
from the previous case is that the rational function 
$Q^{(n)}(w|\zeta )^{\varepsilon}$ contains the factor 
\begin{equation}
\zeta_{2n-1} (xw_{2n-1}-z_{2n})|_{w_{2n-1}=xz_{2n}}=
(x^2 -1)\zeta_{2n-1}z_{2n}, 
\label{eq:Q+-}
\end{equation}
in the present case, while the corresponding factor 
in the previous case is 
\begin{equation}
\zeta_{2n} (xz_{2n-1}-w_{2n})|_{w_{2n}=xz_{2n}}= 
x\zeta_{2n} (z_{2n-1}-z_{2n}). 
\label{eq:Q-+}
\end{equation}
Since (\ref{eq:Q+-}) is equal to $\epsilon$ times 
(\ref{eq:Q-+}) as 
$\zeta_{2n}=\epsilon x^{-1}\zeta_{2n-1}$, the condition 
(\ref{eq:rec-G-comp}) with $\varepsilon_{2n-1}>0$ 
and $\varepsilon_{2n}<0$ follows from the previous 
case. After all, the condition 
(\ref{eq:rec-G-comp}) was componentwisely proved. 

\subsection{Non-trivial theta function}

In subsections 3.1--3.3 we proved Proposition 
\ref{prop:qKZ}. The key point in proving 
(\ref{eq:R-symm-comp}) was the $\overline{R}$-matrix 
symmetry of the rational function 
$Q^{(n)}(w|\zeta )^{\varepsilon}$. We observed that 
the other two conditions (\ref{eq:cyc-comp}) and 
(\ref{eq:rec-G-comp}) hold under the assumption of 
the transformation properties and 
the recursion relation of $\vartheta^{(n)}_\sigma$. 
But actually, you can easily find (\ref{eq:z-sym}), 
(\ref{eq:x-sym}) and (\ref{eq:th-rec}) by 
imposing (\ref{eq:cyc'}) for $\varepsilon_{2n}<0$, 
(\ref{eq:cyc'}) for $\varepsilon_{2n}>0$ and 
(\ref{eq:cyc-comp}), respectively. 

In this way we obtained the integral solutions to 
Smirnov-type master equations for correlation 
functions of the anti-ferromagnetic XXZ model, with the 
freedom of the choice of $\vartheta^{(n)}_\sigma$. 
Now we wish to present an example of 
$\vartheta^{(n)}_\sigma$ satisfying 
all the properties given below (\ref{eq:df-psi}). 
\begin{equation}
\vartheta^{(n)}_\sigma (w|\zeta )
=\Theta_{x^2} \left( -\sigma \prod_{a\in A} w_a^{-1} 
\prod_{j=1}^{2n}\zeta_j \right) \prod_{j=1}^{2n}z_j^{-n} 
\prod_{a,b\in A\atop a<b} w_a^{-1} 
\Theta_{x^2} (w_a /w_b ). 
\label{eq:th-sol}
\end{equation}
You can easily see that (\ref{eq:th-sol}) satisfies 
all the properties of symmetry with respect to 
$\zeta_j$'s, antisymmetry with respect to $w_a$'s, 
(\ref{eq:z-sym}), (\ref{eq:x-sym}) and (\ref{eq:th-rec}). 
Note that this example of the function 
$\vartheta^{(n)}_\sigma (w|\zeta )\prod_{j=1}^{2n}z_j^{n}$ 
coincides with the theta function for the 
anti-ferromagnetic XXZ model form factors \cite{KMQ} 
in the `total spin-$0$' sector, up to a constant. 
We therefore conclude that Smirnov-type integral solutions 
for both form factors and correlation functions hold the 
essentially common integral kernel, and that the only 
difference among both formulae results from the choice 
of the rational functions. 

Taking into account that $x=q^2$ in \cite{JMbk} while we 
set $x=-q$, you can easily confirm that our integral 
formulae with the choice of the function 
$\vartheta^{(n)}_\sigma$ (\ref{eq:th-sol}) 
reproduce the ones obtained in the trace 
construction \cite{JMbk}, up to a trivial constant. 
Here, the following formulae are useful in order to 
compare eq. (8.21) in \cite{JMbk} with our solutions: 
$$
\prod_{a\in A} \zeta_a =
\prod_{j=1}^{2n} \zeta^{(1+\varepsilon_j )/2}, 
$$
and 
$$
\frac{1}{2}\left( 
\Theta_{x^2} (-u)\pm \Theta_{x^2} (u) \right) 
=\left\{ \begin{array}{l} 
\Theta_{x^8} (-x^2 u^2 ), \\
u\Theta_{x^8} (-x^6 u^2 ). \end{array} \right. 
$$

We also notice that there is actually no poles at 
$w_a =x^2 w_b$ ($b<a$) and $w_a =x^{-2}w_b$ ($b>a$) 
when we fix $\vartheta^{(n)}_\sigma$ to (\ref{eq:th-sol}) 
because of the factor $\Theta_{x^2} (w_a /w_b )$. 

\section{Integral formulae for the boundary XXZ model}

\subsection{The Hamiltoninan and the $K$-matrix} 

In this section we consider the XXZ spin chain 
in a half-infinite lattice 
\begin{equation}
H_{bXXZ}=-\frac{1}{2}\sum_{j=1}^\infty 
(\sigma_{j+1}^x \sigma_{j}^x + \sigma_{j+1}^y 
\sigma_{j}^y +\Delta \sigma_{j+1}^z \sigma_{j}^z) 
+h\sigma_{1}^z , 
\label{eq:bXXZ-H}
\end{equation}
where 
\begin{equation}
\Delta=-\frac{x+x^{-1}}{2}, ~~~~
h=\frac{1-x^2}{4x}\frac{1+r}{1-r}. 
\label{eq:df-h}
\end{equation}
We again restrict ourselves to the anti-ferromagnetic 
regime: $0<x<1$. We also restrict the discussion below 
to the case $0<x^2 <|r|<z^{-1}<1$ such that $h>0$. Since 
the quantity $h$ denotes the magnegic field at the boundary 
site, the $\mathbb{Z}_2$-symmetry is broken by the boundary 
term. In what follows we assume that there is only one 
ground state for the anti-ferromagnetic XXZ model with 
a boundary. In the presence of the positive magnetic field 
at the boundary, we should fix the unique ground state 
to $|0\rangle_B$, in the terminology of Ref. \cite{JKKKM}. 

As done in 
the bulk case, the Hamiltonian (\ref{eq:bXXZ-H}) 
can be obtained from the transfer matrix for the 
six vertex model on a half-infinite lattice, by taking 
logarithmic derivative with respect to the spectral 
parameter $\zeta$. The boundary interaction in the 
boundary six vertex model is specified by the following 
diagonal reflection matrix \cite{RE,Skl}
\begin{equation}
K(\zeta )v_\varepsilon =\sum_{\varepsilon'=\pm} 
v_{\varepsilon'} K(\zeta )^{\varepsilon'}_\varepsilon , 
~~~~
K(\zeta )=\frac{1}{f(z;r)}\overline{K}(\zeta ). 
\label{eq:K-mat}
\end{equation}
Here the scalar function $f(z;r)$ is given by 
$$
f(z;r)=f_1 (z;r)f_2 (z), ~~~~
f_1 (z;r)=\frac{\varphi_1 (z; r)}{\varphi_1 (z^{-1}; r)}, 
~~~~ f_2 (z)=\frac{\varphi_2 (z^2 )}{\varphi_2 (z^{-2})}, 
$$
and 
\begin{equation}
\varphi_1 (z; r)=\frac{(x^2 rz;x^4)_{\infty}}
{(x^4 rz;x^4)_{\infty}}, ~~~~
\varphi_2 (z^2 )=\frac{(x^8 z^{2};x^8)_{\infty}}
{(x^6 z^2 ;x^8)_{\infty}}. 
\label{eq:varphi}
\end{equation}
The non-zero entries are 
\begin{equation}
\overline{K}(\zeta )^+_+ =k_+ (z)=
\frac{1-rz}{z-r}, ~~~~
\overline{K}(\zeta )^-_- =k_- (z)=1. 
\label{eq:K-comp}
\end{equation}
We also introduce another $K$-matrix $K' (\zeta )$: 
$$
K' (\zeta )v_\varepsilon =
\sum_{\varepsilon'=\pm} v_{\varepsilon'} 
K(x^{-1}\zeta )^{-\varepsilon'}_{-\varepsilon}. 
$$
The explicit form of $K' (\zeta )$ is given as 
follows: 
\begin{equation}
K' (\zeta )=\frac{1}{f'(z;r)}\overline{K'}(\zeta ), 
~~~~
f'(z;r)=x^{-2}z
\frac{\varphi_1 (z^{-1}; r)}{\varphi_1 (x^{-4}z; r)}
\frac{\varphi_2 (x^2 z^{-1}; r)}{\varphi_2 (x^{-2}z; r)}, 
\label{eq:K*-mat}
\end{equation}
and the non-zero entries are 
\begin{equation}
\overline{K'}(\zeta )^+_+ =k'_+ (z)=
\frac{1}{k_+ (x^{-2}z)}, ~~~~
\overline{K'}(\zeta )^-_- =k'_- (z)=1. 
\label{eq:K*-comp}
\end{equation}

The main properties of the $K$-matrix 
are the reflection (boundary Yang-Baxter) equation
\begin{equation}
K_2 (z_2 )R_{21}(z_1 +z_2 )K_1 (z_1 )R_{12}(z_1 -z_2 )
=
R_{21}(z_1 -z_2 )K_1 (z_1 )R_{12}(z_1 +z_2 )K_2 (z_2 ). 
\label{eq:RE}
\end{equation}
the initial condition 
\begin{equation}
K(1)=I; 
\label{eq:b-ini}
\end{equation}
and the unitarity relation
\begin{equation}
K(\zeta)K(\zeta^{-1})=1; 
\label{eq:b-uni}
\end{equation}
the $\mathbb{Z}_2$-parity 
\begin{equation}
K(-\zeta)=K(\zeta); 
\label{eq:KZ_2}
\end{equation}
and the boundary crossing symmetry
\begin{equation}
K(\zeta)^{\varepsilon'_1}_{\varepsilon_1}
=\sum_{\varepsilon_2 , \varepsilon'_2 =\pm} 
    R(\zeta^2)_{\varepsilon'_2 -\varepsilon'_1}^{
    -\varepsilon_1 \varepsilon_2} 
    K(\epsilon x^{-1}\zeta )^{
    \varepsilon'_2}_{\varepsilon_2} 
    ~~~~ (\epsilon =\pm ). 
\label{eq:b-cross}
\end{equation}
Under this normalization the partition function per 
boundary lattice site is equal to unity in the 
thermodynamic limit \cite{JKKKM}. 

\subsection{Correlation functions and difference equations}

Let us introduce the $V^{\otimes 2n}$-valued 
correlation functions 
\begin{equation}
G^{(n)}_b (\zeta_1,\cdots,\zeta_{2n})=
\sum_{\varepsilon_j =\pm\atop \varepsilon_1 +
\cdots +\varepsilon_{2n}=0} v_{\varepsilon_1} \otimes 
\cdots \otimes v_{\varepsilon_{2n}} 
G^{(n)}_b (\zeta_1,\cdots,\zeta_{2n})
^{\varepsilon_1 \cdots \varepsilon_{2n}}. 
\label{eq:df-bcorr}
\end{equation}
Here, we restrict $G^{(n)}_b (\zeta)$ 
to the `total spin-$0$' subspace of $V^{\otimes 2n}$. 
The subscript $b$ refers to quantities in the boundary 
model. In the bulk model, we considered two correlation 
functions $G^{(n)}_\pm (\zeta)$ for each $n$. On the contrary, 
because of the  $\mathbb{Z}_2$-symmetry breakdown, 
we will consider only one correlation function 
$G^{(n)}_b (\zeta)$ for each $n$ in the boundary model. 
As in the bulk model case, by specializing 
the spectral parameters the correlation 
function (\ref{eq:df-bcorr}) gives the expectation 
value of the local operator as follows: 
\begin{equation}
\langle E^{(1)}_{\varepsilon_1 \varepsilon'_1} 
\cdots E^{(n)}_{\varepsilon_n \varepsilon'_n} 
\rangle_b 
=G^{(n)}_b (\overbrace{x^{-1}\zeta,\cdots,
x^{-1}\zeta}^{n}, \overbrace{\zeta,\cdots,\zeta}^{n})^{
-\varepsilon_{n}\cdots -\varepsilon_{1}\varepsilon'_1 
\cdots \varepsilon'_{n}}. 
\label{eq:b-spec}
\end{equation}

In this section we often use the abbreviations 
$(\tilde{\zeta})=(\zeta_2 , \cdots , \zeta_{2n})$, and 
$(\tilde{\varepsilon})=(\varepsilon_2 , \cdots , 
\varepsilon_{2n})$. For fixed indices $a_1 , 
\cdots , a_n$ such that $a_1 <\cdots <a_n$, we also use 
the abbreviation $(\tilde{w})=(w_{a_2}, \cdots , w_{a_n})$. 
On the basis of the boundary CTM bootstrap 
approach, the correlation functions satisfy 
the following four conditions \cite{JKKMW}: 

\noindent{\it 1. $R$-matrix symmetry}
\begin{equation}
P_{j\,j+1} G_{b}^{(n)} 
(\cdots,\zeta_{j+1},\zeta_j,\cdots) 
\quad =
R_{j\,j+1}(\zeta_j/\zeta_{j+1})G_{b}^{(n)} 
(\cdots,\zeta_j,\zeta_{j+1},\cdots)
\qquad (1\leqslant j\leqslant 2n-1). 
\label{eq:bR-symm} 
\end{equation}

\noindent{\it 2. Reflection properties}
\begin{equation}
K_{2n} (\zeta_{2n}) G_{b}^{(n)} (\zeta', \zeta_{2n})
=G_{b}^{(n)} (\zeta', \zeta^{-1}_{2n}); 
\label{eq:RE-1}
\end{equation}
and 
\begin{equation}
K'_{1}(\zeta_1 )
G_{b}^{(n)} ( \zeta_{1}^{-1}, \tilde{\zeta})
=G_{b}^{(n)} (x^{-2}\zeta_{1}, \tilde{\zeta}). 
\label{eq:RE-2}
\end{equation}

\noindent{\it 3. Normalization}
\begin{equation}
\begin{array}{rcl}
G_{b}^{(n)} (\zeta'',\zeta_{2n-1}, \zeta_{2n})|_{
\zeta_{2n}=\epsilon x^{-1}\zeta_{2n-1}} &=&
G_{b}^{(n-1)} (\zeta'')\otimes u_\epsilon ~~~~
(\epsilon =\pm). 
\end{array}
\label{eq:rec-bG}
\end{equation}

Since the $R$-matrix and $K$-matrix preserves the 
`total spin' of the correlation function, we can 
regard $G_{b}^{(n)}(\zeta)$ as the element of the 
`total spin-$0$' subspace, when we consider the equations 
(\ref{eq:bR-symm}--\ref{eq:rec-bG}). Note that the first 
three equations imply the boundary analogue of 
the quantum KZ difference equation: 
\begin{equation}
\begin{array}{rcl}
T_j G^{(n)}_b (\zeta ) 
&=& R_{j j-1}(x^{-2}\zeta_j /\zeta_{j-1}) 
\cdots R_{j 1}(x^{-2}\zeta_j /\zeta_{1}) 
\hat{K}_j(\zeta_j ) \\
&\times&
R_{1 j}(\zeta_1 \zeta_{j}) \cdots
R_{j-1 j}(\zeta_{j-1}\zeta_j ) 
R_{j+1 j}(\zeta_{j+1}\zeta_j ) \cdots
R_{2n j}(\zeta_{2n}\zeta_j ) \\
&\times& K_j (\zeta_j ) 
R_{j 2n}(\zeta_j /\zeta_{2n})  \cdots
R_{j j+1}(\zeta_j /\zeta_{j+1})
G^{(n)}_b (\zeta), 
\label{eq:N-diff}
\end{array}
\end{equation}
where $T_j$ is the shift operator defined by 
(\ref{eq:T-shift}). 

\subsection{Integral formulae} 

Set 
\begin{equation}
 G_{b}^{(n)}(\zeta)
=c'_n \prod_{1\leqslant j< k \leqslant 2n} \zeta_j g(z_j/z_k)
g(z_j z_k) \prod_{j=1}^{2n} \varphi_1 (z_j ;r) g_b (z_j )
\times \overline{G}_{b}^{(n)}(\zeta). 
\label{eq:bG-bar}
\end{equation}
Here $c'_n$ is a constant which will be determined below, 
and the scalar function $g_b (z)$ has the properties 
\begin{equation}
g_b (z)=g_b (x^{-2}z^{-1}), ~~~~ 
f_2 (z)=\frac{g_b (z)}{g_b (z^{-1})}. 
\label{eq:gb-prop}
\end{equation}
The explicit form of $g_b (z)$ is as follows: 
\begin{equation}
g_b (z)=\frac{
(x^{10}z^2;x^4,x^8)_{\infty}(x^6 z^{-2};x^4,x^8)_{\infty}}
{(x^{12}z^2;x^4,x^8)_{\infty}(x^8 z^{-2};x^4,x^8)_{\infty}}. 
\label{eq:df-gb}
\end{equation}

Thanks to (\ref{eq:g-prop}) and (\ref{eq:gb-prop}) the 
first three equations (\ref{eq:bR-symm}--\ref{eq:RE-2}) are 
rephrased in terms of 
$\overline{G}_{b}^{(n)}(\zeta)$ 
and $\overline{R}(\zeta)$ as follows 
\begin{equation}
\overline{G}_{b}^{(n)} 
(\cdots,\zeta_{j+1},\zeta_j,\cdots)^{\cdots 
\varepsilon_{j+1}\varepsilon_j\cdots}
=\sum_{\varepsilon'_j , \varepsilon'_{j+1}=\pm} 
\overline{R}(\zeta_j/\zeta_{j+1})^{
\varepsilon_j , \varepsilon_{j+1}}_{
\varepsilon'_j , \varepsilon'_{j+1}}
\overline{G}_{b}^{(n)} 
(\cdots,\zeta_j,\zeta_{j+1},\cdots)^{
\varepsilon'_j , \varepsilon'_{j+1}}, 
\label{eq:bR-symm'} 
\end{equation}
\begin{equation}
k_{\varepsilon_{2n}}(\zeta_{2n})
\overline{G}_{b}^{(n)} (\zeta', \zeta_{2n})^{
\varepsilon'\,\varepsilon_{2n}}
=\overline{G}_{b}^{(n)} 
(\zeta',\zeta_{2n}^{-1})^{\varepsilon'\,\varepsilon_{2n}}, 
\label{eq:RE1'}
\end{equation}
\begin{equation}
(x^2 z_1^{-1})^{2n}k'_{\varepsilon_{1}}(\zeta_{1})
\overline{G}_{b}^{(n)} (\zeta_{1}^{-1},\tilde{\zeta})^{
\varepsilon_1\,\tilde{\varepsilon}}
=\overline{G}_{b}^{(n)} 
(x^{-2}\zeta_{1},\tilde{\zeta})^{
\varepsilon_1\,\tilde{\varepsilon}}, 
\label{eq:RE2'}
\end{equation}

Let us define the 
following $V^{\otimes 2n}$-valued meromorphic function 
\begin{equation}
Q_b^{(n)}(w|\zeta )=\sum_{\varepsilon_j =\pm\atop 
\varepsilon_1 +\cdots +\varepsilon_{2n}=0} 
v_{\varepsilon_1}\otimes \cdots \otimes 
v_{\varepsilon_{2n}} 
Q_b^{(n)}(w|\zeta )^{\varepsilon}, 
\label{eq:vec-Qb}
\end{equation}
where 
\begin{equation}
Q_b^{(n)}(w|\zeta )^{\varepsilon} 
=Q^{(n)}(w|\zeta )^{\varepsilon} q_b^{(n)}(w|\zeta ), 
~~~~
q_b^{(n)}(w|\zeta )=
\prod_{a\in A} \prod_{j=1}^{2n} (xz_j -w_a^{-1}). 
\label{eq:df-Qb}
\end{equation}
Since the factor $q_b^{(n)}(w|\zeta )$ is symmetric 
with respect to $\zeta_j$'s and $w_a$'s, respectively, 
the $R$-matrix symmetry (\ref{eq:bR-symm'}) is also valid 
in the boundary case. 

We wish to find integral formulae of the form 
\begin{equation}
\overline{G}_{b}^{(n)} (\zeta)=
\prod_{a\in A}\oint_{C_a} 
\dfrac{dw_a}{2\pi iw_a} 
\Psi_{b}^{(n)} (w|\zeta )
Q_b^{(n)}(w|\zeta). 
\label{eq:Gb-form}
\end{equation}
Here, the kernel has the form 
\begin{equation}
\Psi^{(n)}_b (w| \zeta )=
\vartheta^{(n)}_b (w | \zeta )
\prod_{a\in A}\prod_{j=1}^{2n} 
\psi_b (w_{a},z_j ),
\label{eq:df-Psib}
\end{equation}
where
\begin{equation}
\psi_b (w, z)=\psi \left( \frac{w}{z} \right) 
\psi \left( wz \right) =\frac{1}{(xw/z;x^4)_{\infty}
(xz/w;x^4)_{\infty}(xwz;x^4)_{\infty}
(x/zw;x^4)_{\infty}}. 
\label{eq:df-psib}
\end{equation}
Note that $\psi_b (w, z)=\psi_b (w, z^{-1})=
\psi_b (w^{-1}, z)$. 
For the function 
$\vartheta^{(n)}_b (w|\zeta )$ 
we assume that
\begin{itemize}
\item
it is anti-symmetric and holomorphic
in the $w_a \in \mathbb{C} \backslash \{x^{-1}r^{-1}\}$,
 
\item
it is symmetric and meromorphic
in the $\zeta_j\in \mathbb{C} \backslash \{0\}$,
 
\item
it has the four transformation properties 
\begin{eqnarray}
\displaystyle\frac
{\vartheta^{(n)}_b (w |\zeta', \zeta^{-1}_{2n})}
{\vartheta^{(n)}_b (w | \zeta )}&=& z_{2n}^{2n}, 
\label{eq:z^-1}\\
\displaystyle \frac
{\vartheta^{(n)}_b (w', w^{-1}_{a_n}| \zeta )}
{\vartheta^{(n)}_b (w | \zeta )}
&=& -\dfrac{w_{a_n}^{-1}-rx}{w_{a_n}-rx}
\displaystyle\prod_{a\in A\atop a\neq a_n} 
\frac{x^{-1}w_{a_n}^{-1}-xw_a}{x^{-1}w_{a_n}-xw_a}, 
\label{eq:w^-1} \\
\displaystyle\frac
{\vartheta^{(n)}_b (w |x^{-2}\zeta_1,\tilde{\zeta})}
{\vartheta^{(n)}_b (w | \zeta )}&=& x^{4n}, 
\label{eq:bz-sym}\\
\displaystyle \frac
{\vartheta^{(n)}_b (x^{-4}w_{a_1},\tilde{w}| \zeta )}
{\vartheta^{(n)}_b (w | \zeta )}
&=& x^{-12n}w_{a_1}^{4n} 
\dfrac{1-xrw_{a_1}}{1-x^{-3}rw_{a_1}} 
\displaystyle\prod_{a\in A\atop a\neq a_1} 
\frac{x^{-1}w_{a_1}^{-1} -xw_a}
{x^{-1}w_{a_1}^{-1} -x^{-3}w_a}. 
\label{eq:bx-sym}
\end{eqnarray}
\item it satisfies the following recursion relation
\begin{equation}
\begin{array}{l}
\displaystyle \frac{\vartheta_{b}^{(n)}
(w', x^{-1}z_{2n-1} |\zeta'',\zeta_{2n-1}, 
\epsilon x^{-1} \zeta_{2n-1})}
{\vartheta_{b}^{(n-1)}(w' |\zeta'')}=
\displaystyle x^{2n}z_{2n-1}^{-2n}\prod_{j=1}^{2n-2} 
z_j^{-1} 
\\[6mm] 
\qquad\qquad\times
\dfrac{\Theta_{x^4}(x^{-2}z_{2n-1}^2)}{1-rz_{2n-1}} 
\displaystyle\prod_{a\in A\atop a\neq a_{2n}} 
\frac{\Theta_{x^2}(x w_{a}/z_{2n-1})
\Theta_{x^2}(x^{-1}z_{2n-1}w_{a})}{z_{2n-1}^{-1}-xw_a}. 
\end{array}
\label{eq:bth-rec}
\end{equation}
Here we also fix the constant $c'_n$ as follows: 
\begin{equation}
c'_n =\frac{x^{2n}(x^2 ; x^4 )_\infty^{n}
(x^2 ; x^4 , x^4 )_\infty^{2n}}{
(x^2 ; x^2 )_\infty^{n^2}
(x^8 ; x^4 , x^4 )_\infty^{2n}}. 
\label{eq:c'_n}
\end{equation}
\end{itemize}
The function $\vartheta_{b}^{(n)}(w|\zeta )$
is otherwise arbitrary, and the choice of 
$\vartheta_{b}^{(n)}(w|\zeta )$ 
corresponds to that of solutions to 
(\ref{eq:bR-symm}--\ref{eq:rec-bG}). 

{}From (\ref{eq:vec-Qb}, \ref{eq:w^-1}, 
\ref{eq:bx-sym}, \ref{eq:bth-rec}), 
we find that the integrand may have poles at 
$$
w_a =\left\{ \begin{array}{ll}
x^{\pm (1+4k)}z_j & 
1\leqslant j\leqslant 2n$, $k\in \mathbb{Z}_{\geqslant 0}, \\
x^{1+4k}z_j^{-1} & 
1\leqslant j\leqslant 2n$, $k\in \mathbb{Z}_{\geqslant 0}, \\
x^{-5-4k}z_j^{-1} & 
1\leqslant j\leqslant 2n$, $k\in \mathbb{Z}_{\geqslant 0}, \\
x^{-1}r^{-1}. & \end{array} \right. 
$$
We choose the integration contour $C_a$ with 
respect to $w_a$ ($a\in A$) such that $C_a$ is along 
a simple closed curve oriented anti-clockwise, and 
encircles the points $x^{1+4k}z_j^{\pm 1}$ 
($1\leqslant j \leqslant 2n , k\in \mathbb{Z}_{\geqslant 0}$), 
but not $x^{-1-4k} z_j$, $x^{-5-4k}z_j^{-1}$ 
($1\leqslant j \leqslant 2n, k\in \mathbb{Z}_{\geqslant 0}$) 
nor $x^{-1} r^{-1}$. We further fix the contour $C_a$ 
such that $x^{-1}z_j^{-1}$ lie outside $C_a$, as in the 
bulk model case. 

\subsection{Main propositions}

We are now in a position to state the proposition in 
the boundary anti-ferromagnetic XXZ model case: 

\begin{prop} Assume the properties of 
the function $\vartheta^{(n)}_b$ 
below (\ref{eq:df-psib}) and the integration contour 
$C_a$ below (\ref{eq:c'_n}). Then the integral formulae 
(\ref{eq:df-bcorr}, \ref{eq:bG-bar}, \ref{eq:Gb-form}) 
with (\ref{eq:varphi}, \ref{eq:df-gb}, \ref{eq:vec-Qb}, 
\ref{eq:df-Q}, \ref{eq:df-Qb}, \ref{eq:df-Psib}, 
\ref{eq:df-psib}) solves the four equations 
(\ref{eq:bR-symm}--\ref{eq:rec-bG}). 
\label{prop:b-qKZ}
\end{prop}

A proof will be given in Appendix A. 

The following example of $\vartheta^{(n)}_\sigma$ satisfies 
all the properties given below (\ref{eq:df-psib}): 
\begin{equation}
\vartheta^{(n)}_\sigma (w|\zeta )
=\prod_{j=1}^{2n}z_j^{-n} 
\prod_{a\in A} \frac{\Theta_{x^4} (w_a^{2})}{1-xrw_a}
\prod_{a,b\in A\atop a<b} \frac{w_b 
\Theta_{x^2} (w_a /w_b )\Theta_{x^2} (w_a w_b )}
{x^{-1}-xw_a w_b}. 
\label{eq:bth-sol}
\end{equation}
Taking into account that our integral variables $w_a$'s 
correspond to $q^3 w_a$ in \cite{JKKKM}, 
you can easily confirm that our integral 
formulae with the choice of the function 
$\vartheta^{(n)}_b$ (\ref{eq:bth-sol}) 
reproduce the ones obtained in the trace 
construction \cite{JKKKM}, up to a trivial constant. 

\section{Concluding remarks} 

In this paper we have constructed the correlation 
functions of both the bulk and boundary XXZ model in the 
antiferromagnetic regime. This was done by directly solving 
Smirnov-type master equations; i.e., the $R$-matrix symmetry, 
cyclicity conditions (resp. reflection properties) and 
the normalization conditions for the bulk (resp. boundary) 
case. Precisely speaking, we made the Ansatz that the 
correlation functions of the bulk (resp. boundary) XXZ model 
are expressed in terms of integral transform of the rational 
functions $Q^{(n)}_\sigma(w|\zeta )$ 
(resp. $Q^{(n)}_b (w|\zeta )$) with the integral 
kernels $\Psi^{(n)}_\sigma(w|\zeta )$ (resp. 
$\Psi^{(n)}_b (w|\zeta )$). 

The integral solutions to Smirnov-type master 
equations for the bulk case were given by 
(\ref{eq:df-corr}, \ref{eq:G-bar}, \ref{eq:G-form}) 
with (\ref{eq:df-g}, \ref{eq:vec-Q}, \ref{eq:df-Q}, 
\ref{eq:df-Psi}, \ref{eq:df-psi}), where 
the function $\vartheta^{(n)}_\sigma (w|\zeta )$ 
satisfies (\ref{eq:z-sym}--\ref{eq:th-rec}). Those for 
the boundary case were given by 
(\ref{eq:df-bcorr}, \ref{eq:bG-bar}, \ref{eq:Gb-form}) 
with (\ref{eq:varphi}, \ref{eq:df-gb}, \ref{eq:vec-Qb}, 
\ref{eq:df-Q}, \ref{eq:df-Qb}, \ref{eq:df-Psib}, 
\ref{eq:df-psib}), where the function 
$\vartheta^{(n)}_b (w|\zeta )$ satisfies 
(\ref{eq:z^-1}--\ref{eq:bth-rec}). Our integral formulae 
with appropriate choice of the functions 
$\vartheta^{(n)}_\sigma(w|\zeta )$ and 
$\vartheta^{(n)}_b (w|\zeta )$ reproduce the 
known results \cite{CORR,JKKKM}. The explicit choice was 
given in (\ref{eq:th-sol}) (resp. (\ref{eq:bth-sol})) 
for the bulk (resp. boundary) case. 

In comparison with the form factor integral formulae 
\cite{JKMQ,KMQ,affine}, the corresponding rational function 
has the determinant structure and consequently different 
from $Q^{(n)}_\sigma(w|\zeta )$ obtained in the present 
paper. On the other hand, the integral kernels are essentially 
the same. Now we have a natural question whether the integral 
formulae for form factors of the boundary XXZ model can be 
written by using the integral kernel $\Psi^{(n)}_b (w|\zeta )$. 
We wish to address this problem in a subsequent paper. 

\section*{Acknowledgements}
The author would like to thank A. Nakayashiki for 
valuable discussion. He would also like to thank 
M. Kashiwara and T. Miwa for the invitation to 
the Workshop {\it MATHPHYS ODYSSEY 2001 --- 
Integrable Models and Beyond} held in 
Okayama and Kyoto, where this work was started. 

\appendix

\section{Appendix A. A proof of Proposition 
\ref{prop:b-qKZ}}

The $R$-matrix symmetry (\ref{eq:bR-symm}) can be 
shown by repeating the similar arguments as proving 
(\ref{eq:R-symm}). Straightforward calculations shows 
that the normalization condition (\ref{eq:rec-bG}) 
follows from (\ref{eq:bth-rec}, \ref{eq:c'_n}). 
Here the formulae (\ref{eq:lim_z}--\ref{eq:ppQ_a}) 
are again useful. In order to show the reflection 
properties (\ref{eq:RE-1}, \ref{eq:RE-2}), the following 
observation is very relevant. 
The recursion relation (\ref{eq:bth-rec}) indicates 
that the function $\vartheta^{(n)}_b$ contains the 
factor 
$$
\prod_{a\in A} \frac{\Theta_{x^4} (w_a^{2})}{1-xrw_a}. 
$$
Thus we find that 
\begin{equation}
\mbox{$\vartheta^{(n)}_b(w|\zeta )$ has zeros at $w_a =
x^{2n}$ ($n\in\mathbb{Z}$), \; and has a pole at 
$w_a =x^{-1}r^{-1}$. }
\label{eq:bth-0pole}
\end{equation}

Let us prove (\ref{eq:RE-1}) by showing (\ref{eq:RE1'}). 
First consider the case $\varepsilon_{2n}<0$. By comparing 
the integrands of both sides, the desired equality follows 
from (\ref{eq:z^-1}) and the equality 
$$
\vartheta^{(n)}_b(w|\zeta', \zeta_{2n})\prod_{a\in A} 
(xw_a -z_{2n})(xz_{2n}-w_a^{-1})=
\vartheta^{(n)}_b(w|\zeta', \zeta_{2n}^{-1})\prod_{a\in A} 
(xw_a -z_{2n}^{-1})(xz_{2n}^{-1}-w_a^{-1}). 
$$
When $\varepsilon_{2n}>0$ the relation (\ref{eq:RE1'}) 
is equivalent to 
$$
\frac{1-rz_{2n}}{z_{2n}-r}
\overline{G}_{b}^{(n)} (\zeta', \zeta_{2n})^{
\varepsilon'\,+}
=\overline{G}_{b}^{(n)} 
(\zeta',\zeta_{2n}^{-1})^{\varepsilon'\,+}. 
$$
The difference between both sides are equal to
\begin{equation}
\frac{1}{z_{2n}}\prod_{a\in A}\oint_{C_a} 
\dfrac{dw_a}{2\pi iw_a} 
\Psi_{b}^{(n)} (w|\zeta )
Q_b^{(n)}(w|\zeta)\left( 
\frac{1-rz_{2n}}{1-rz_{2n}^{-1}}-
\frac{xw_{2n}-z_{2n}}{xw_{2n}-z_{2n}^{-1}} \right). 
\label{eq:RE-1+}
\end{equation}
Since the integral (\ref{eq:RE-1+}) vanishes, 
the relation (\ref{eq:RE1'}) with $\varepsilon_{2n}>0$ 
holds. In order to see this, let us consider the change of 
the integral variable $w_{2n}\mapsto w_{2n}^{-1}$. Under 
this transform, the integral contour $C_{2n}$ is invariant 
while the integrand of (\ref{eq:RE-1+}) negates the sign. 
The latter follows from (\ref{eq:z^-1}) and the equality 
$$
\sum_{s=\pm} \vartheta^{(n)}_b 
(w',w_{2n}^s|\zeta' ,\zeta_{2n})(w_{2n}^{-s}-xr)
\prod_{a\in A\atop a\neq 2n} (x^{-1}w_{2n}^s -xw_a )^{-1}
=0. 
$$
Thus the one of the reflection properties (\ref{eq:RE-1}) 
is proved. 

Finally, let us prove another reflection property 
(\ref{eq:RE-2}) by showing (\ref{eq:RE2'}). 
First consider the case $\varepsilon_{1}<0$. Since the 
function $Q_b^{(n)}(w|\zeta)$ contains the factor 
$(xz_1 -w_a )(xz_1 -w_a^{-1})$ for any $a\in A$, there 
is no pole at $w_a =(xz_1 )^{\pm 1}$ in the integrand of 
$\overline{G}_{b}^{(n)} (\zeta)^{-\,\tilde{\varepsilon}}$. 
When the RHS of (\ref{eq:RE2'}) with $\varepsilon_{1}<0$ 
is analytically continued from $\zeta_{1}$ to 
$x^{-2}\zeta_{1}$, the poles of the integrands 
$$
\cdots\cdots, \, x^{13} z_{1}^{\pm 1}, \, 
x^9 z_{1}^{\pm 1}, \, x^5 z_{1}^{\pm 1}, \, 
x z_{1}^{-1}, \, x^{-1}z_{1}, 
\, x^{-5}z_{1}^{\pm 1}, \, x^{-9}z_{1}^{\pm 1}, \, 
x^{-13}z_{1}^{\pm 1}, \, \cdots\cdots , 
$$ 
move to the points 
$$
\cdots\cdots, \, x^{13} z_{1}^{\pm 1}, \, 
x^9 z_{1}^{\pm 1}, \, x^5 z_{1}^{\pm 1}, \, 
x z_{1}, \, x^{-1}z_{1}^{-1}, 
\, x^{-5}z_{1}^{\pm 1}, \, x^{-9}z_{1}^{\pm 1}, \, 
x^{-13}z_{1}^{\pm 1}, \, \cdots\cdots . 
$$ 
Thus the integral contour $C_a$ for any $a\in A$ is invariant 
even if we make the analytical continuation 
$\zeta_1 \rightarrow x^{-2}\zeta_1$. Furthermore, 
by using (\ref{eq:bz-sym}), the desired equality 
follows from that of the integrands. 

When $\varepsilon_{1}>0$ the LHS of (\ref{eq:RE2'}) 
times $(x^{-2}z_1)^{2n-1}$ reduces to 
\begin{equation}
\prod_{a\in A}\oint_{C_a} 
\dfrac{dw_a}{2\pi iw_a} 
\Psi_{b}^{(n)} (w|\zeta_1^{-1},\tilde{\zeta})
Q_b^{(n)}(w|\zeta_1^{-1},\tilde{\zeta})
\frac{1-x^2 rz_{1}^{-1}}{1-x^{-2}rz_{1}}. 
\label{eq:RE-2+L}
\end{equation}
Concerning the RHS of (\ref{eq:RE2'}), the integral 
contour $C_a$ for $a\in A\backslash\{ 1\}$ is again 
invariant from the same reason as the previous case. 
As for the the integral with respect to $w_1$, we 
have to keep in mind that the integrand of 
$\overline{G}_{b}^{(n)} (\zeta)^{+\,\tilde{\varepsilon}}$ 
has a pole at $w_1 =xz_1$. Thus the RHS of (\ref{eq:RE2'}) 
times $(x^{-2}z_1)^{2n-1}$ reduces to 
\begin{equation}
\prod_{a\in A\backslash\{ 1\}}\oint_{C_a} 
\dfrac{dw_a}{2\pi iw_a} 
\left( \oint_{C_1} +2\pi i
\mbox{Res}_{w_1 =x^{-3}z_1} \right) 
\dfrac{dw_1}{2\pi iw_1} 
\Psi_{b}^{(n)} (w|\zeta_1^{-1},\tilde{\zeta})
Q_b^{(n)}(w|\zeta_1^{-1},\tilde{\zeta})
\frac{xz_{1}^{-1}-w_1}{x^{-3}z_{1}-w_1}. 
\label{eq:RE-2+R}
\end{equation}
Let us denote (\ref{eq:RE-2+L}) by $L$ and 
(\ref{eq:RE-2+R}) by $R$, respectively. Furthermore, 
we will divide $R$ into the sum of three parts as follows. 
In (\ref{eq:RE-2+R}) we denote the integral part 
with respect to $w_1$ by $R_1$; and the residue part 
by $R_2 -R_3$, the difference of two contour integrals. 
For that purpose, we introduce the function 
$$
R(w)=
\frac{x^{-3}z_1 -x^{-4}w_1^{-1}}{w_1 -x^{-4}w_1^{-1}}
\frac{1-xrw_1}{1-x^{-2}rz_1}, 
$$
as we have done in \cite{JKMQ}. 
The integrand of (\ref{eq:RE-2+R}) has poles at 
$w_1 =x^{-3}z_1, xz_1^{-1}, x^{-1}r^{-1}$ in the 
annulus between the closed lines $C_1$ and $x^{-4}C_1$. 
Here, the closed line $x^{-4}C_1$ denotes the one 
obtained from $C_1$ by similarity transformation with the 
ratio being $x^{-4}$. Since 
$R(x^{-3}z_1 )=1$ and $R(xz_1^{-1})=R(x^{-1}r^{-1})=0$, 
the residue at $w_1 =x^{-3}z_1$ can be evaluated as follows: 
\begin{equation}
\prod_{a\in A\backslash\{ 1\}}\oint_{C_a} 
\dfrac{dw_a}{2\pi iw_a} 
\left( \oint_{x^{-4}C_1} - \oint_{C_1} \right) 
\dfrac{dw_1}{2\pi iw_1} 
\Psi_{b}^{(n)} (w|\zeta_1^{-1},\tilde{\zeta})
Q_b^{(n)}(w|\zeta_1^{-1},\tilde{\zeta})
\frac{xz_{1}^{-1}-w_1}{x^{-3}z_{1}-w_1}R(w_1 ). 
\label{eq:RE-2Res}
\end{equation}
Let us denote the first and the second term of 
(\ref{eq:RE-2Res}) by $R_2$ and $-R_3$, respectively. 
The function $R(w_1 )$ has a pole at $w_1 =x^{-2}$, 
however, we notice that the integrand of 
(\ref{eq:RE-2Res}) is regular at this point because 
of (\ref{eq:bth-0pole}). 
The integral $R_2$ can be rewritten in terms of the 
integral along the contour $C_1$ by changing the integral 
variable $w_1 \mapsto x^{-4}w_1^{-1}$ as follows: 
$$
R_2 =\prod_{a\in A\backslash\{ 1\}}\oint_{C_a} 
\dfrac{dw_a}{2\pi iw_a} 
\oint_{C_1} 
\dfrac{dw_1}{2\pi iw_1} 
\Psi_{b}^{(n)} (w|\zeta_1^{-1},\tilde{\zeta})
Q_b^{(n)}(w|\zeta_1^{-1},\tilde{\zeta})
\frac{xz_{1}^{-1}-x^{-4}w_1^{-1}}{w_1 -x^{-4}w_1^{-1}}
\frac{1-xrw_1}{1-x^{-2}rz_1}. 
$$
Thus we obtain 
$$
L-R_1 =R_2 -R_3 =
\prod_{a\in A}\oint_{C_a} \dfrac{dw_a}{2\pi iw_a} 
\Psi_{b}^{(n)} (w|\zeta_1^{-1},\tilde{\zeta})
Q_b^{(n)}(w|\zeta_1^{-1},\tilde{\zeta})
\frac{x^{-3}z_{1}-xz_1^{-1}}{x^{-3}z_{1}-w_1}
\frac{1-xrw_1}{1-x^{-2}rz_1}, 
$$
which implies (\ref{eq:RE2'}) with $\varepsilon_{1}>0$. 
Thus Proposition \ref{prop:b-qKZ} was proved.

\end{document}